\documentclass[conference]{IEEEtran}
\IEEEoverridecommandlockouts
\usepackage{bm}

\usepackage{multirow}
\usepackage{tabularx}
\usepackage{longtable}
\usepackage{threeparttable}
\usepackage[export]{adjustbox}
\usepackage{multicol}
\usepackage{booktabs}
\usepackage{hhline}

\usepackage{cite}
\usepackage{amsmath,amssymb,amsfonts}
\usepackage{algorithmic}
\usepackage{graphicx}
\usepackage{textcomp}
\usepackage{xcolor}

\def\BibTeX{{\rm B\kern-.05em{\sc i\kern-.025em b}\kern-.08em
    T\kern-.1667em\lower.7ex\hbox{E}\kern-.125emX}}
\begin{document}

\title{Mask-CTC-based Encoder Pre-training for Streaming End-to-End Speech Recognition}

\author{
\IEEEauthorblockN{
    Huaibo Zhao$^{\star}$, 
    Yosuke Higuchi$^{\star}$, 
    Yusuke Kida$^{\dagger}$,
    Tetsuji Ogawa$^{\star}$,
    Tetsunori Kobayashi$^{\star}$ 
}
\IEEEauthorblockA{
\textit{$^{\star}$Department of Communications and Computer Engineering, Waseda University, Tokyo, Japan}\\
\textit{$^{\dagger}$LINE Corporation, Tokyo, Japan}\\
\texttt{zhao@pcl.cs.waseda.ac.jp}}}


\maketitle

\begin{abstract}
Achieving high accuracy with low latency has always been a challenge in streaming end-to-end automatic speech recognition (ASR) systems.
By attending to more future contexts,
a streaming ASR model achieves higher accuracy but results in larger latency, which hurts the streaming performance.
In the Mask-CTC framework, an encoder network is trained to learn the feature representation that anticipates long-term contexts, which is desirable for streaming ASR.
Mask-CTC-based encoder pre-training has been shown beneficial in achieving low latency and high accuracy for triggered attention-based ASR.
However, the effectiveness of this method has not been demonstrated for various model architectures, nor has it been verified that the encoder has the expected look-ahead capability to reduce latency.
This study, therefore, examines the effectiveness of Mask-CTC-based pre-training for models with different architectures, such as Transformer-Transducer and contextual block streaming ASR.
We also discuss the effect of the proposed pre-training method on obtaining accurate output spike timings,
which contributes to the latency reduction in streaming ASR.
\end{abstract}


\section{Introduction}
\label{sec:intro}

In recent years, deep learning has become the core technology of automatic speech recognition (ASR)~\cite{hinton2012deep, Chorowski2015AttentionBasedMF}.
End-to-end ASR further integrates the traditional separated components (i.e., acoustic, pronunciation, and language models) into a single deep neural network, significantly contributing to the simplicity of ASR developments~\cite{Graves2014TowardsES,Chorowski2015AttentionBasedMF,Chan2016ListenAA}.
End-to-end ASR models can be realized in various approaches, including connectionist temporal classification (CTC)~\cite{graves2006connectionist}, Transducer~\cite{Graves2012SequenceTW}, and attention-based encoder-decoder~\cite{Chorowski2015AttentionBasedMF,Chan2016ListenAA}.
These end-to-end ASR approaches have greatly benefited from the adoption of Transformer~\cite{vaswani2017attention,Zhang2020TransformerTA,Chen2021DevelopingRS,Dong2018SpeechTransformerAN,Karita2019ImprovingTE}, enabling a model to capture global contexts using the self-attention mechanism.

Streaming properties (i.e., real-time processing) are of vital importance in the applications of ASR systems.
With the superior performance of Transformer, many efforts have been devoted to making Transformer-based ASR models streaming.
Triggered attention-based ASR~\cite{moritz2020streaming} obtains alignment information from CTC and realizes frame-synchronous decoding according to the CTC spike timings.
Meanwhile, contextual block streaming ASR (CBS-ASR)~\cite{Tsunoo2021StreamingTA} splits the input into blocks (chunks), and streaming encoder feature extraction is conducted on each block with the contexts inherited from the previous blocks.
A block boundary detection algorithm is applied to detect the index boundary in each block, which enables block-synchronous beam search decoding.
Apart from the above streaming models based on attention-based encoder-decoder, the Transducer-based model can be naturally applied to streaming ASR.
Transducer trains a model to align the output of the acoustic encoder with the output of the label encoder, enabling frame-synchronous decoding and making it a suitable framework for streaming ASR~\cite{Zhang2020TransformerTA}.
Transformer-Transducer (Transformer-T)~\cite{Chen2021DevelopingRS} adopts Transformer for the acoustic encoder, where the chunk-wise attention mask limits the look-ahead range of the self-attention layer to ensure streaming properties.


For streaming ASR models in general, performance degradation occurs when the look-ahead range is limited from global to local, suppressing the advantage of the Transformer architecture (i.e., processing with long-range contexts).
Consequently, longer look-ahead ranges are often required to provide adequate future contexts, leading to the growth of latency requirements. 
Therefore, capturing long-term contexts within short look-ahead ranges is essential for building successful streaming ASR systems.

One approach to realizing such a property is to utilize the Mask-CTC~\cite{higuchi2020mask, higuchi2021improved} framework.
With conditional masked language model (CMLM)~\cite{Devlin2019BERTPO, ghazvininejad2019mask} and CTC multi-task training, Mask-CTC trains an encoder network to extract acoustic feature representation that contributes to capturing long-term output dependencies by anticipating future contexts.
Its capability of learning context-rich bi-directional representations has been validated in the spoken language understanding task~\cite{Li2023NonAutoregressiveEA, Meeus2022BidirectionalRF}.
In our previous work~\cite{Zhao2021AnIO}, we conducted supervised pre-training with the Mask-CTC objective on the encoder and CTC modules of the triggered attention-based model, which has shown effective results for improving accuracy while reducing latency.
In this work,
we aim to further examine the effectiveness of Mask-CTC-based pre-training for streaming models with various architectures.

In addition, a detailed perspective on the latency can be analyzed by focusing on the timing of output spikes.
In streaming ASR, the timing of output spikes is generally delayed due to the lack of long-term context information.
We expect that the proposed pre-training based on Mask-CTC will introduce the look-ahead capability to the encoder, thereby enabling accurate prediction of the timing of output spikes.
This property contributes to the early determination of recognition results, which is essential for streaming applications (e.g., a system that interacts with a user in real-time).


This study, therefore, attempts to demonstrate the effectiveness of Mask-CTC-based pre-training for streaming models with different architectures, including Transformer-T and CBS-ASR,
and to discuss the contribution of the proposed pre-training to latency reduction.




The rest of this paper is organized as follows.
Section~\ref{sec:techniques} overviews the streaming ASR models and the Mask-CTC model.
Section~\ref{sec:proposal} describes the proposed pre-training approach for constructing low latency and high recognition accuracy streaming ASR models.
In Section~\ref{sec:exp}, we demonstrate the effectiveness of Mask-CTC-based pre-training through experiments and discuss the latency reduction effect with output spike timing measurements.
Finally, Section~\ref{sec:conclusion} concludes this paper.

\section{Background}
\label{sec:techniques}


In this section, we introduce two types of streaming ASR approaches that we study in this work: Transformer-Transducer~\cite{Chen2021DevelopingRS} and contextual block streaming ASR~\cite{Tsunoo2021StreamingTA}. 
We also describe Mask-CTC~\cite{higuchi2020mask}, which is the key to our proposed method.

\subsection{Transformer-Transducer}
\label{ssec:tt}
A Transducer-based ASR model contains three components: acoustic encoder, label encoder, and joint network. 
Given a streaming input to a current time index $t$,
the output probability of the $u$-th token is calculated as follows:
\begin{align}
&\bm{\mathrm{h}}_{t}^{\mathsf{AE}} = \mathrm{AcousticEncoder}(\bm{\mathrm{x}}_{1:t})\label{eq:1},
\\
&\bm{\mathrm{h}}_{u-1}^{\mathsf{LE}} = \mathrm{LabelEncoder}(y_{1:u-1})\label{eq:2},
\\
&\bm{\mathrm{h}} = \mathrm{Tanh}(\mathrm{Linear}(\bm{\mathrm{h}}_{t}^{\mathsf{AE}})+\mathrm{Linear}(\bm{\mathrm{h}}_{u-1}^{\mathsf{LE}}))\label{eq:3},
\\
&P(y_{u}|y_{1:u-1}, \bm{\mathrm{x}}_{1:t}) = \mathrm{SoftMax}(\bm{\mathrm{h}})\label{eq:4}.
\end{align}
First, the acoustic encoder embeds the input sequence $\bm{\mathrm{x}}_{1:t}$ into vector $\bm{\mathrm{h}}_{t}^{\mathsf{AE}}$ (Eq.~\eqref{eq:1}).
Meanwhile, the label encoder generates $\bm{\mathrm{h}}_{u-1}^{\mathsf{LE}}$ from the previous output token sequence $y_{1:u-1}$ (Eq.~\eqref{eq:2}).
The two outputs are then sent to the joint network, projected to the same dimension, and added up (Eq.~\eqref{eq:3}).
Finally, the output probabilities against tokens in a vocabulary $\mathcal{V}$ are calculated based on the previous result (Eq.~\eqref{eq:4}).
The Transducer framework predicts the current symbol for each input frame based on the past output tokens, which naturally introduces streaming fashion into decoding.

Various neural network types can be applied to implement the acoustic and label encoders~\cite{Hochreiter1997LongSM,Graves2012SequenceTW,He2019StreamingES}.
In the work of ~\cite{Chen2021DevelopingRS}, Transformer~\cite{vaswani2017attention} is applied to the acoustic encoder to achieve high accuracy and LSTM~\cite{Hochreiter1997LongSM} for the label encoder in consideration of the model size control.
Chunk-wise attention masks are applied to the self-attention layers of the Transformer acoustic encoder to enable streaming feature extraction.
This architecture is referred to as a Transformer-Transducer (Transformer-T).

\subsection{Contextual block streaming ASR}
\label{ssec:cbs}
Contextual block streaming ASR (CBS-ASR)~\cite{Tsunoo2021StreamingTA} introduces streaming properties to attention-based encoder-decoder models.
For streaming feature extraction in the encoder, CBS-ASR utilizes block processing with a context inheritance mechanism proposed in~\cite{Tsunoo2019TransformerAW}.
The speech input is segmented into blocks containing past, central, and future frames with the numbers of $N_l$, $N_c$, and $N_r$.
The input blocks are passed on to the encoder, where the central frames are utilized for the output with local contexts provided by the past and future frames as well as the global contexts provided by a context embedding vector inherited from the previous block. 
Streaming decoding is achieved by a block boundary detection (BBD) algorithm~\cite{Tsunoo2021StreamingTA},
which takes end-of-sentence prediction or token repetition as stopping criteria from detecting the index boundaries on-the-fly and enables the beam search synchronous to the encoded blocks.
The streaming processing in CBS-ASR
is calculated as follows:
\begin{align}
&H_{b}, \bm{\mathrm{c}}_{b} = \mathrm{BlockEncoder}(Z_{b},\bm{\mathrm{c}}_{b-1}),
\label{eq:5}
\\
&\alpha(y_{0:i},H_{1:B})\approx\sum_{b=1}^{B} \sum_{j=I_{b-1}+1}^{I_b} \log p(y_i| y_{0:j-1},H_{1:b}).
\label{eq:6}
\end{align}
Eq.~\eqref{eq:5} represents the streaming encoding of the $b$-th input sequence $Z_{b}$, where $|Z_{b}| = N_l+N_c+N_r$.
The encoded acoustic features $H_{b}$ is obtained from $Z_{b}$ and the contextual vector from the previous block $\bm{\mathrm{c}}_{b-1}$.
Eq.~\eqref{eq:6} represents the score of the partial hypothesis $y_{0:i}$ during streaming beam search decoding, where $y_{0}$ is the start-of-sequence token.
$I_b$ denotes the index boundary of the $b$-th input block derived from the BBD algorithm.

\subsection{Mask-CTC}
\label{secsec:maskctc}

The Mask-CTC framework~\cite{higuchi2020mask} aims to learn feature representations suitable for anticipation of future contexts.
Mask-CTC trains an encoder-decoder model with the joint CMLM~\cite{Devlin2019BERTPO} and CTC objectives.
During training, tokens in the ground truth are randomly masked, and the masked tokens are predicted based on contextual information captured by the encoder and other unmasked output tokens.
For the input $X$ and observed tokens $Y_{\text{obs}}$,
the output probabilities of the masked tokens $Y_{\text{mask}}$ are computed as follows:
\begin{equation}
    P_{\text{cmlm}}(Y_{\text{mask}}|Y_{\text{obs}}, X)  = \prod\limits_{y\in Y_{\text{mask}}} P_{\text{cmlm}}(y|Y_{\text{obs}}, X),
\end{equation}
where $Y_{\mathrm{obs}}$ is $Y \setminus Y_{\mathrm{mask}}$. 
Based on the CMLM mask prediction of the decoder,
the encoder network of Mask-CTC is trained to consider the long-term bidirectional dependencies between output tokens,
which enables it to generate acoustic feature representations that anticipate future information.

Such properties are desirable in streaming ASR as it allows the model to capture more future contexts with a limited look-ahead range.
In such a way,
the Mask-CTC framework can be a potential solution for improving streaming ASR,
enhancing a model to achieve high accuracy while keeping low latency.

\begin{figure}[tb]
\begin{minipage}[b]{0.95\linewidth}
\centering
\centerline{\includegraphics[width=\linewidth]{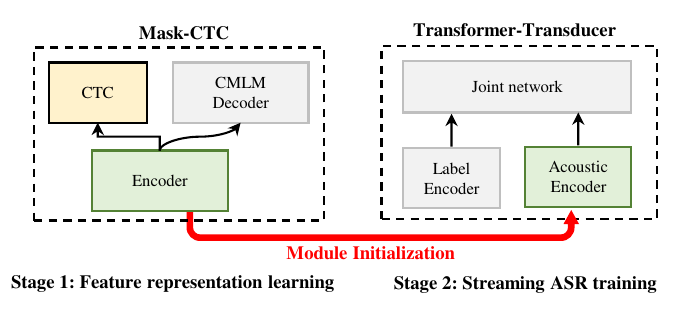}}
\caption{Illustration of Mask-CTC-based pre-training using Transformer-Transducer model. In stage 1, encoder is trained with Mask-CTC framework. In stage 2, Transformer-Transducer model is initialized with pre-trained encoder and fine-tuned with streaming objective.}
\label{fig:proposal}
\end{minipage}
\end{figure}

\section{Mask-CTC-based pre-training method}
\label{sec:proposal}

We present a simple and general Mask-CTC-based pre-training method for achieving high-accuracy and low-latency streaming ASR.
Specifically, this paper aims to demonstrate the effectiveness of the Mask-CTC pre-training regardless of model architectures and discusses whether such pre-training can extract features suitable for anticipation as intended, focusing on the alignment of the output tokens.

As different end-to-end streaming ASR models, we focus on Transformer-T (see Section~\ref{ssec:tt}) and CBS-ASR (see Section~\ref{ssec:cbs}), which cover both Transducer and encoder-decoder model architectures.
For both models, the adoption of Transformer has realized high recognition accuracy in their non-streaming baselines.
However, when applied to streaming scenarios, the look-ahead ranges of self-attention layers are limited from global to local.
This leads to an inevitable performance drop by degrading the Transformer's capability to capture long-range contexts, which limits applications where low latency is a top priority for recognition.

To remedy such an effect, we need the feature representation for the input sequence that considers long-term contextual dependencies and anticipates future information, which corresponds to the properties of the Mask-CTC encoder network as described in Section~\ref{secsec:maskctc}.
To introduce the desirable properties of the Mask-CTC model into the streaming ASR, we propose a simple two-step training method as follows, which is also described in Fig.~\ref{fig:proposal}:
\begin{itemize}

\item\textbf{Stage 1 (Feature representation learning):}
The Mask-CTC model is pre-trained to obtain an encoder network that can consider long-term dependencies and anticipate future information.

\item\textbf{Stage 2 (Streaming ASR training):}
The pre-trained Mask-CTC model is exploited to initialize the streaming ASR models.
For Transformer-T, the acoustic encoder with the chunk-wise attention is initialized with the Mask-CTC encoder. 
For CBS-ASR, both the Mask-CTC encoder and CTC networks are used to initialize the corresponding components. 
\end{itemize}

With the two-step training method above, we expect to inherit the characteristics of Mask-CTC to a streaming ASR model to capture long-term contextual information and reduce the latency dependency.

\section{Experiments}
\label{sec:exp}
Speech recognition experiments were conducted to examine the effectiveness of the Mask-CTC-based pre-training method using ESPnet2~\cite{Watanabe2018ESPnetES, Boyer2021ASO}.
We also investigated the essential effect of the proposed pre-training method by studying the output token alignments of the streaming ASR models.

\subsection{Datasets}

The models were trained and evaluated using the Wall Street Journal (WSJ)~\cite{paul1992design} dataset, which contains 81h English utterances of read articles from the newspaper and the TED-LIUM2 (TED2)~\cite{Rousseau2014EnhancingTT} dataset, which contains 207h English spontaneous speech.
For the output tokens,
we used SentencePiece~\cite{kudo2018sentencepiece} to construct a 80 subword vocabulary for WSJ and a 500 subword vocabulary for TED2, respectively.
For robust model training, 
we applied SpecAugment~\cite{Park2019SpecAugmentAS} to the input data.

\subsection{Experimental setup}
For the Transformer-T model, the acoustic encoder was implemented with 12 Transformer encoder layers and a single LSTM layer for the label encoder. 
For streaming feature extraction, a chunk-wise attention mask was implemented and applied to the encoder layers as in \cite{Chen2021DevelopingRS}.
The latency value was calculated as the product of the maximum look-ahead range (i.e., $\text{chunk size} - 1$) and a frame rate of 40ms. 

For WSJ experiments, the CBS-ASR model consisted of 6 Conformer encoder layers~\cite{Gulati2020ConformerCT} and 6 Transformer decoder layers.
The input block settings followed $N_l$ as eight, $N_c$ as four, and $N_r$ varying from 0 to 6.
The latency for CBS-ASR was
calculated as the product of the maximum look-ahead range in the block
(i.e., $N_c + N_r -1$) and a frame rate of 40ms.
For TED2 experiments, the CBS-ASR model consisted of 12 Conformer encoder layers~\cite{Gulati2020ConformerCT} and 6 Transformer decoder layers. The $N_r$ was set to 6.

For the pre-trained Mask-CTC model, the encoder was constructed with the identical setting as the target streaming model. The CMLM decoder was built with six Transformer decoder layers.
All the models were trained by 150 epochs, and the final models were obtained by averaging the snapshots of the ten epochs of the minimal loss for Transformer-T and the best accuracy for CBS-ASR.
For decoding, a beam search was conducted with a beam size of ten for all.
We used the word error rate (WER) for measuring the ASR performance.
\begin{table}[htbp]
\caption{Word error rates on WSJ dataset. 
}
\begin{center}
\label{table1}
\begin{tabular}{ l c c c c }
    \toprule
    & & & \multicolumn{2}{c}{\textbf{WER} [\%] ($\downarrow$)} \\
    \cmidrule(l{0.5em}r{0.5em}){4-5}
    \textbf{Model} & \textbf{Latency} [ms] & \textbf{Initialization} & eval92 & dev93 \\
    \midrule
    \textbf{Baseline} &  &  &  &  \\
    \midrule
    \multirow{4}{*}[-3pt]{Transformer-T} & 120 & \multirow{3}{*}[0pt]{Random} & 19.5 & 23.3\\
    & 160 &  & 16.8 & 20.9 \\
    & 200 &  & \textbf{15.1} & \textbf{18.9} \\
    \cmidrule{2-5}
    & $\infty$ & Random & 14.7 & 17.3\\
    \midrule
    \multirow{4}{*}[-3pt]{CBS-ASR} & 200 & \multirow{3}{*}[0pt]{Random}  & 14.4 & 18.1 \\
    & 280 &  & 13.2 & 16.2 \\
    & 360 &  & \textbf{12.9} & \textbf{16.1} \\
    \cmidrule{2-5}
    & 1240 & Random & 11.2 & 14.2\\
    \midrule\midrule
    \textbf{Enhanced} &  &  &  &  \\
    \midrule
    \multirow{3}{*}[0pt]{Transformer-T} & 120 & \multirow{3}{*}[0pt]{Mask-CTC}  & 16.6 & 20.8 \\
    & 160 &  & 15.0 & 19.0 \\
    & 200 &  & \textbf{14.8} & \textbf{18.5} \\
    \midrule
    \multirow{3}{*}[0pt]{CBS-ASR} & 200 & \multirow{3}{*}[0pt]{Mask-CTC} & 13.5 & 17.2\\
    & 280 &  & 12.9 & \textbf{16.0} \\
    & 360 &  & \textbf{12.2} & 16.1 \\
    \bottomrule
\end{tabular}

\end{center}
\end{table}
\begin{table}[htbp]
\caption{Word error rate on TED2 dataset.}
\begin{center}
\label{table2}

\begin{tabular}{ l c c c }
    \toprule
    \textbf{Model} & \textbf{Latency} [ms] & \textbf{Initialization} & \textbf{WER} [\%] ($\downarrow$)  \\
    \midrule
    \textbf{Baseline} &  &  &   \\
    \midrule
    \multirow{2}{*}[-3pt]{CBS-ASR} & 280 & Random & 11.3  \\
    \cmidrule{2-4}
    & 1240 & Random & 9.8 \\
    \midrule\midrule
    \textbf{Enhanced} &  &  &    \\
    \midrule
    CBS-ASR & 280 & Mask-CTC & 11.1 \\
    \bottomrule
\end{tabular}

\end{center}
\end{table}


\subsection{Experimental results}


For both the Transformer-T and CBS-ASR systems, the performances of the following models are compared.
\begin{itemize}
\item \textbf{Baseline}~\cite{Chen2021DevelopingRS, Tsunoo2021StreamingTA, Moritz2019TriggeredAF}:
Existing streaming ASR models, including Transformer-T and CBS-ASR.
The parameters for all the components were randomly initialized.

\item \textbf{Enhanced}:
Streaming ASR models with Mask-CTC-based pre-training.
Components of the streaming ASR were initialized with pre-trained Mask-CTC modules.
For Transformer-T, the acoustic encoder was initialized with the Mask-CTC encoder.
For CBS-ASR, both encoder and CTC modules were initialized with corresponding Mask-CTC modules.
\end{itemize}


The experimental results of Transformer-T and CBS-ASR are summarized in Table~\ref{table1} and Table~\ref{table2}.
Non-streaming Transformer-T and CBS-ASR with 1240ms latency were used as lower bounds in the experiments.

The results on WSJ show that for both Transformer-T and CBS-ASR, the enhanced models outperformed the baseline models by achieving lower WERs under all latency settings, suggesting the accuracy enhancements introduced by the Mask-CTC-based pre-training method.
For WSJ dataset, 40ms and 80ms latency reductions were reached for Transformer-T and CBS-ASR, respectively, while achieving better or equal recognition accuracy than the baseline models.
For instance, the enhanced Transformer-T with 120ms latency achieved lower WERs (16.6\% for eval92 and 20.8\% for dev93) than the WERs of the baseline with 160ms latency (16.8\% for eval92 and 20.9\% for dev93).
Such results demonstrated that our method contributed to the construction of streaming ASR models with low latency and high accuracy.
For TED2 dataset, the enhanced CBS-ASR model also achieved 0.2 percentage point of WER reduction compared to the baseline model, 
which proves the general effectiveness of the proposed method regardless of the dataset.
The results for systems with different architectures, such as Transducer and Encoder-Decoder, also demonstrated that the Mask-CTC-based pre-training was effective regardless of the model architecture.
\begin{figure}[t]
\begin{minipage}[b]{0.95\linewidth}
\centering
\centerline{\includegraphics[width=0.99\linewidth]{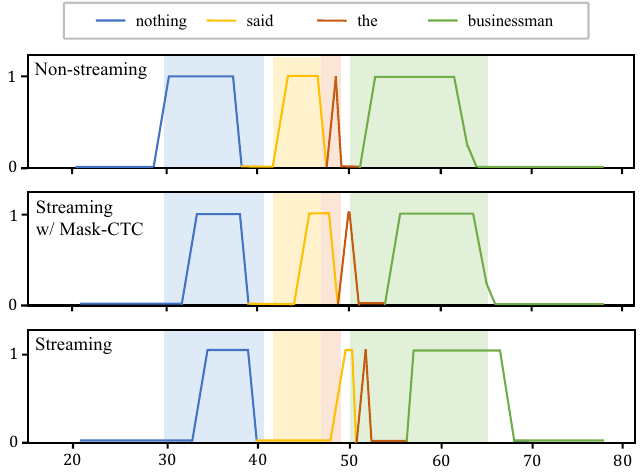}}
\caption{Output token alignments of non-streaming and streaming Transformer-Transducer models. 
}
\label{fig:spike}
\end{minipage}
\end{figure}

\subsection{Analysis of output token alignments}


The work of~\cite{Inaguma2021AlignmentKD} argued that the streaming model attends to shift the token boundaries to the future side to obtain more contextual information,
which results in delay of the posterior probability spikes for the output tokens compared to non-streaming models.
In contrast, if the encoder network learns the feature representations that anticipate future information, the output tokens can be confirmed earlier and the token boundary shifting issue should be remedied in some instances.
Therefore, 
we measured the delay of the spike occurrences in streaming models by comparing them to the alignments obtained from a non-streaming model.
The delay is expected to be reduced with the Mask-CTC-based pre-training method.

We conducted measurements on the dev93 validation set of WSJ.
We used the baseline and enhanced models with 200ms latency settings for Transformer-T and compared their alignments with a non-streaming Transformer-T model.
The alignments were obtained from the output of the joint network. 
For CBS-ASR, the latency was also set to 200ms, and we compared the output token boundaries between the baseline and enhanced models. 
The ASR alignments were obtained from the CTC predictions of CBS-ASR in the same manner as~\cite{Inaguma2021AlignmentKD} and the reference alignments were obtained with the Montreal Forced Aligner~\cite{McAuliffe2017MontrealFA}.
Figure \ref{fig:spike} illustrates one example of output token alignments given by Transformer-T.
Here, the color in the background represents the reference alignment to the speech input.
The non-streaming ASR (top) managed to predict accurate token alignments.
However, the baseline streaming ASR (bottom) showed a significant delay in the alignments, indicating token boundary shifting due to the lack of contexts.
Meanwhile, our enhanced streaming ASR (middle), with a Mask-CTC-based pre-trained encoder network, 
largely improved the alignments of the streaming ASR.
We calculated the average output delay reduction across the dev93 validation set for both Transformer-T and CBS-ASR. 
For Transformer-T, the spike output delay was reduced by 44ms, and for CBS-ASR, 46ms.
Such results help us to understand the knowledge learned from the Mask-CTC-based pre-training method and the reason for the latency reduction capability.

\section{Conclusion}
\label{sec:conclusion}

In this study, an attempt was made to demonstrate the effectiveness of Mask-CTC-based pre-training for achieving low latency and high accuracy in streaming speech recognition.
Experimental results showed the effectiveness of the method on various model architectures, 
including Transformer-Transducer and contextual block streaming ASR.
Furthermore, by studying the output spike timings of the streaming models, 
we discovered that more precise alignments of the input and output sequences are learnt by the pre-training,
which contributes to the latency reduction in streaming ASR.

\bibliographystyle{IEEEbib}
\bibliography{strings,refs}


\end{document}